# Optimization of Test Case Generation using Genetic Algorithm (GA)

Ahmed Mateen
Department of Computer
Science,
University of Agriculture
Faisalabad, Pakistan

Marriam Nazir
Department of Computer
Science,
University of Agriculture
Faisalabad, Pakistan

Salman Afsar Awan, PhD
Department of Computer
Science,
University of Agriculture
Faisalabad, Pakistan

## ABSTRACT

Testing provides means pertaining to assuring software performance. The total aim of software industry is actually to make a certain start associated with high quality software for the end user. However, associated with software testing has quite a few underlying concerns, which are very important and need to pay attention on these issues. These issues are effectively generating, prioritization of test cases, etc. These issues can be overcome by paying attention and focus. Solitary of the greatest Problems in the software testing area is usually how to acquire a great proper set associated with cases to confirm software. Some other strategies and also methodologies are proposed pertaining to shipping care of most of these issues. Genetic Algorithm (GA) belongs to evolutionary algorithms. Evolutionary algorithms have a significant role in the automatic test generation and many researchers are focusing on it. In this study explored software testing related issues by using the GA approach. In addition to right after applying some analysis, better solution produced, that is feasible and reliable. The particular research presents the implementation of GAs because of its generation of optimized test cases. Along these lines, this paper gives proficient system for the optimization of test case generation using genetic algorithm.

## 1. INTRODUCTION
Computer software assessment is one of the majorities of labor strenuous as well as pricey period with the software program improvement lifetime routine. Computer software assessment consists of the test circumstance, age group as well as test suite optimization that includes a strong impact on the particular usefulness as well as productivity connected with software program assessment. In the last number of ages, there was an energetic investigation to automate the task connected with the test circumstance, age group, however the tribes are already confined by the dimension and the complexity connected with the software program. The high quality software must satisfy the user requirements and customer demands. Satisfaction of customer has always been important because it indicates that business people have to manage and improve their product. Once testing involving software is actually a good important process connected with assessing the software to help distinguish it is quality. It is an important area of the software engineering. Modern software systems are extremely reliable in addition to correct.

Represented an optimization tool where their aim is to find a problem solution to a given problem. Based on inheritance, natural selection, mutation, and sexual reproduction, they try to give after many generations the optimal solution in a finite time [1]. Testing techniques are test case design method. Test cases are developed using various testing techniques to achieve more effective testing of application [2]. Genetic algorithms are generally as outlined by evolutionary ideas connected with natural menu in addition to genetics. Genetic algorithms solve the current problems step via step as well as provide and then generation [3].

Testing is a basic activity of the item headway handles, and robotic test period adds to reduce cost and time trials. The ideal test suites are conceived by the strategy for examining measurements. Testing is the most essential stage in the item change life cycle. The testing stage is the last channel for all oversights of rejection and commission. Testing writing computer programs are altogether more eccentric than rehearsing a framework to check whether it works. Each study, audit, survey, walks around, social event code sees, all is when in doubt a kind of test. The more fruitful that can make early is static attempting, the less issues involved in the dynamic periods of testing. IT has shown again and again that the earlier a defect recognized and cleared, the lower the additional change cost associated with ousting the mix-up. Plan for testing begin when each item thing is portrayed. All evolutionary algorithms similar to Genetic Algorithm acquire near optimal solution.

The essential objective of testing is to exhibit that the item thing at any rate, meets a course of action of pre-set up affirmation criteria under a suggested set of environmental circumstances. There are two sections of this objective. The essential fragment is to show that the requirements point of interest from which the item was laid out is correct. The second portion is to show that the setup and coding precisely responds to the necessities. Precision suggests that limit, execution, and timing necessities match affirmation criteria.

The existing Genetic Algorithm (GA) starts receiving from building a principal population connected with folks, each manifested via randomly earned genotype. The present health of people is normally examined in a lot of problems-dependent technique, and also the GA make an attempt that can help progress highly suit persons from the first population. Algorithms were created via Ruben Netherlands about the sixties and also were produced from the Netherlands as well as the students as well as acquaintances about the University about the state of Michigan. Holland's unique objective feel, not necessarily every single child style algorithms every single child remedy were produced difficulties, but alternatives that can help basically examine the occurrence about adaptation Every bit as that happens with mother nature and also every single child generate methods that the mechanisms concerning pure adaptation might be imported with pass [4].

The genetic algorithms tend to base towards the tip of genetics and evolution. Holland proposed GA as being a heuristic method as outlined by "Survival of any fittest". A genetic algorithm is actually a good evolutionary algorithm, which my partner and solve optimization problem. Then receive approximate merchandise in order to optimize ailments within GA. The genetic algorithm loops greater than a great iteration method to make the current population evolve [5].

The genetic algorithm is actually a stochastic search technique that is to base on the idea of the menu of any fittest chromosome. Throughout genetic algorithm, population of chromosomes represented as coming from other codes including binary, real number, permutation etc. genetic operators (i.e. Selection, crossover, mutation) is usually applied for the chromosome in order to obtain extra fittest chromosome. The fitness of the chromosome is defined by a great proper objective function. As has been a class associated with the stochastic method, genetic algorithm is another coming from a random search. Whilst genetic algorithm executes a good multidimensional search by maintaining a population involving potential user, random actions consisting of the combination regarding iterative search steps as well as simple random search steps probably acquire a solution for an issued problem [6].

## 2. PREVIOUS WORK

The testing, including execution of a system by a few arrangements of test information and contrast the outcomes and expected results are called programming testing. Producing of robotic test information is extremely troublesome errand in item arranged project since legacy, strategy superseding, polymorphism and formats demonstrate numerous coupling abnormalities because of element conduct of articles. Class is the fundamental building square of article situated programming. Customary testing like basic testing, practical determination based testing and heuristics testing are utilized for it. Auxiliary, testing is vital on the grounds that it's found the bugs in codes by control stream testing, way scope testing, information stream testing. Utilitarian testing meets the necessities and detail of programming. Heuristics testing system test the theoretical classes. Programmed test information is created in article situated programming from generally code-based procedure and model-based or plan based strategy [7].

A good technique is usually proposed to abilities in order to prioritize check situation scenarios via figuring out the personal vital way groups utilizing ancestral algorithm. The individual check situation scenarios usually are derived from the UML task diagram along with point out chart diagram. The current screening performance is really optimized for using people ancestral algorithm for the check facts. The personal points movement metric is really implemented of the execute intended for computing the data movement intricacy relevant to every node from the task diagram together with point out chart diagram. Whether the software Requirements alter, people software requirements to assist always modified and also require retesting of the software [8].

Test information usually are generated with most of these, a way they may conduct every record at the least once. Genetic Algorithms applied to path testing no matter whether the target paths are generally clearly defined, as well as a good suitable fitness performs concerning this goal is usually built. GA involves the amount of factors, which happen to commanded to end up being collection consequently sizing of any inhabitants likewise pushed to end up being collected. Human population sizing offers incredible result for that GA swiftness to uncover the best possible answer [9].

Automated generation connected with test cases to evaluate software product is usually much needed. Equally the item probably decreases the night out as well as costs associated with the testing process. Whilst guidebook testing is actually very night out consuming and also costly, Software solutions are right now turning towards the automated testing tools as well as techniques. Inside the paper, explored how Genetic Algorithm, improves quality as well as reliability of any software through bringing in optimized test cases. Three basic steps responsible regarding GA are (1) selection of initial chromosomes through the population, (2) performing crossover from exchanging the facts between choosing chromosomes and also (3) performing a mutation operation towards the selected genes. The actual paper furthermore provides the automated software testing architecture [10].

Genetic Algorithm (GA), Particle Swarm Optimization (PSO) in addition to a good hybrid Genetic Particles warm system algorithm (HGPSTA) has been designed for fitness operate which is according to the dominance relation between only two nodes. The fitness performs based on the criteria connected with facts flow coverage. This is accepted throughout the research, depending to the dominance relation between nodes involving details flow graph. The main goal associated with research, in order to combine the power involving two algorithms and also PSO. This proves their power and effectiveness on the solving the current testing disorders [11].

Examination data creation is basically versions process regarding determining an incredible collection involving data that meet versions standards collection created for testing. The large amount of research is carried out as a result of several scientists as well as developed several test information devices similar to arbitrary test data devices, symbolic test data devices along with powerful test facts devices. The particular document utilized the set analyze of the test scenario creation depends on the hereditary formula and in addition builds test situations [12].

GA inspired from Darwin's theory information about evolution. Genetic algorithm functionalities rule involving menu to evolve a set connected with goods along with looking for a great optimum solution. Genetic algorithms simulate the survival of the fittest among men and women in excess of generations regarding solving a good problem throughout nature competition among persons results on the fittest persons dominating the weaker ones. GAs uses several operations like selection, crossover and mutation [13].

The software testing remains the existing primary course of action, accustomed to obtain consumers' confidence with the software besides categorized the personal check instances making use of stratified sampling. Anatomical formula (GA) offers a general-purpose search system, where benefits guidelines with regards to pure progress. Within the paper, genetic algorithm considered pertaining to producing check selection coming from the individually distinct established linked to check instances. That's why; this is a big challenge. The kind of utilizing a great genetic algorithm on the inside software testing creates a new optimized check selection. The advancement connected with tactics that moreover help types automation linked with software testing certain outcome using substantial financial savings [14].

The use of meta-heuristic worldwide lookup approaches for software program test info age group is the particular concentrate connected with research workers in recent times. Numerous fresh methods as well as cross strategies have also been planned to handle the challenge more effectively [15].

## 3. PROPOSED WORK

The real purpose of this work is to provide better optimization approach, which is introduced in the test case by using Genetic Algorithm. Optimization approach accommodated different scale projects to inspect. This research also provides a survey to determine better quality testing process within the time.

For testing many techniques used by other researchers to be able to accomplish the effective testing process, pertaining to guarantee associated with much better performance along with quality. But those have some limitations. In order to address these issues, there is a study that analyzes how test cases can be optimized and gives best solution.

Evolutionary tests generally are a very good growing methodology associated with routinely bringing in high quality analyzes information. The actual evolutionary algorithms are now put on inside many correct living problems. GA is solitary these kinds of evolutionary criteria. GA offers emerged to become a practical, powerful marketing process together with seeks procedure. A fantastic GA is usually a seek criteria. This is prompted with the approach character evolves variety using an organic food selection of just about any fittest men and women. Any analyses of unique computer software test methods probably executed, to confirm in which GA is successfully used.

### 3.1 Experimental design

In this research first, generated random test cases. Applied mutates testing to check it. If satisfied, then stop.

**Optimized algorithm**

1. Inject the mutant in the program.

2. Generate random test cases.

3. Find the mutation score with the formula, mutation score= (number of mutations found) / (total number of mutants).

4. If the mutant score is satisfactory (Maximum) stop, otherwise go to step 5.

5. Refine the test case using mutation score. Test case, having mutation score 20% or less drops them.

6. Apply Genetic Algorithm Operations on remaining test cases to produce new experiments. Go to step 3.

Algorithm for ab. Where a and b are positive numbers.

1. Power(a, b)
2. If(a= =1)
3. Return 1
4. If(b= =1)
5. Return a
6. P=1, i=1
7. While(i<=b)
8. {P=P*a
9. i++}
10. return P

Inject four mutants in this program

Now algorithm look like this.

1. Power(a, b)
2. If(a=1)
3. Return 1
4. If(b=1)
5. Return a
6. P=1, i=1
7. While(i<b)
8. {P=P+a
9. i++}
10. return P

According to the optimized algorithm flow as shown in the figure 1, found the number of mutants by applying this algorithm. First of all, some mutants added in a program. Mutants are some errors, which encouraged finding the optimized test case. Then evaluate the test case performance and the first step was initialized. After that considered, all mutants not found. The question was what is the number of mutants found? For this purpose next step is the evaluation of mutant numbers found. If found umber is less than the minimum number of mutants then it means test case failed to find errors. On the other hand, the condition is no, then calculated the fitness by finding the exact numbers of mutants. On the next step if mutant numbers found more than 50%, then it displays best fit and if found mutants are less than 50% mutants then it displays mutant number's value. If the total number of mutants found in the 2nd step, then no more steps needed to find errors, just end. Thus the test case provided the optimal solution by finding all the errors.

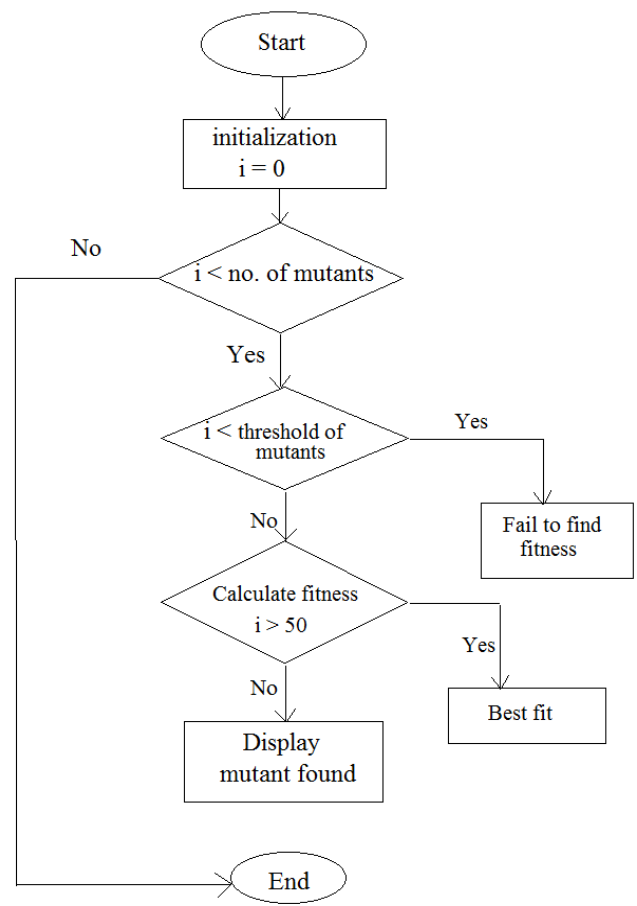

**Fig 1: Optimized algorithm flow**

There has to characterize an approach to decide the nature of every experiment as an issue of reviewing. It is seen that the understudies are a solid propensity to rehash botches. On the off chance that they missed one class, they likely did that for a large portion of the experiments. At that point saw designs among a significant number of the understudies, on why fizzled. Each demonstrating an absence of understanding why the relating learning is vital for experiment outline:

A. Regulated depiction of experiment execution

B. Make particulars of substantial and invalid inputs

C. Understanding directions/level of points of interest

D. Elaborate experiment creation, and not just utilizing the most evident experiment or info

E. Comprehend the motivation behind the framework and current level and connection of testing

F. Characterize a reasonable beginning position for the experiment

G. Understanding test techniques and how to apply them

H. Suppositions, e.g. concerning accuracy and culmination of particulars

I. Experiment assessment (ventures to take to make an unmistakable correlation with expected result ought to be clear)

J. Tidy up after an experiment, repeatability

These factors are playing very important role in test cases. By analyzing it finds that elaborated factors having different level of influence on test cases. All these factors have different impact ratio and these entire ratio described in fig. 2.

**Table 1. Factors and impact ratio**

| Factors | Impact (%) |
|---|---|
| Regulated depiction | 55% |
| Valid & invalid inputs | 38% |
| Good Detail level | 44% |
| Variation in test cases | 75% |
| Understand system's framework | 50% |
| Clear beginning position | 60% |
| Test design techniques better | 80% |
| Accurate suppositions | 70% |
| Test case assessment | 50% |
| Tidy up after execution | 78% |

## A. Regulated depiction of experiment execution

An experiment system is regularly portrayed as an arrangement of activities, with an orderly depiction of activities (and possibly additionally halfway reactions). To make the execution way exceptionally characterized, the experiment, frequently portrayed in little and extremely itemized strides, the very same route as composing code.

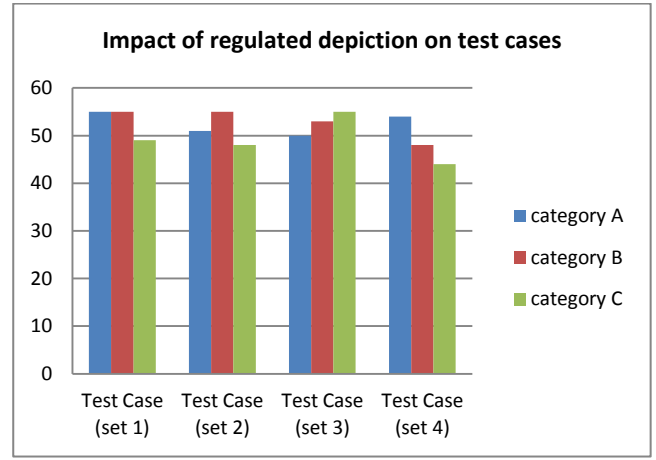

**Fig 2: Impact of regulated depiction**

This graph illustrates regulated depiction impact on different test cases. Here are four sets of test cases, in which maximum impact of regulated depiction is 55%. Each set contains three test cases; in the 1st set impact of the regulated depiction of test cases are respectively 55, 55 and 49. In the 2nd set, impact of the regulated depiction is respectively 51, 55 and 48. In the $3^{rd}$ set, impact is respectively 50, 53 and 55. While in the 4th set impact is respectively 54, 48 and 44. Approximately, 55 % of the experiments gave deficient data and subtle element to give steps that are unambiguously trailed by other human or required personality jumps that are included while making a system for execution.

## B. Substantial and invalid inputs

Portraying data classes in the test/affirmation, determination enhances the usage of different test diagram methodologies, and the data decision of the test. This is a proficient approach to catch the whole info area, furthermore get ready for a progression of test outline strategies. In the meantime, acquaint variables with speak to inputs, in this manner preparing for test computerization. At first hard to characterize what an information is in this setting, subsequent to tapping on a predefined menu-thing can at another reflection level be seen as data.

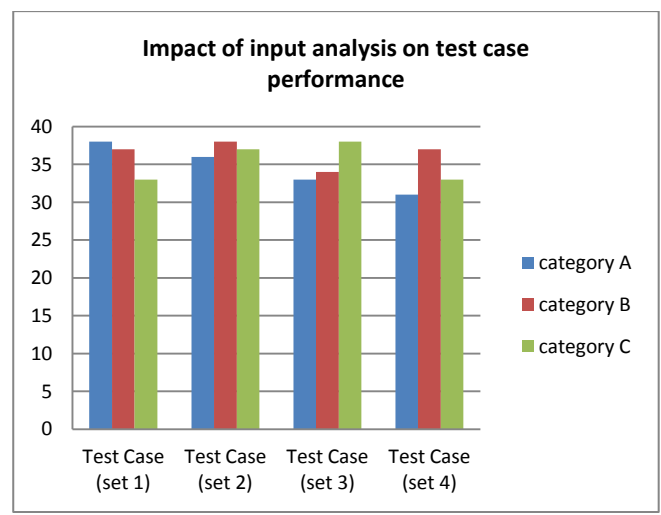

**Fig 3: Impact of input analysis**

This graph describes the impact of input analysis on different test cases. Here are four sets of test cases, in which maximum impact of input analysis is 38%. Each set contains three test cases; in the 1st set impact of the input analysis of test cases are respectively 38, 37 and 33. In the 2nd set, impact of the input analysis is respectively 36, 38 and 37. In the 3$^{rd}$ set, impact is respectively 33, 34 and 38. While in the 4th set impact is respectively 31, 37 and 33. This turned out to be extremely troublesome for the analyzers, and just 62 % were even close to the thought proposed with information examination. The understudies were by and large preferred at giving legitimate contribution over making an examination on invalid information. None succeeded to get a handle on the whole informational space. In this way its impact is 38%.

## C. Understanding directions/level of points of interest

A tester must ready to construe precisely what is implied, and in a nutty gritty level take after directions and give enough data, so that the experiment is unambiguous and can be rehashed by some other tester.

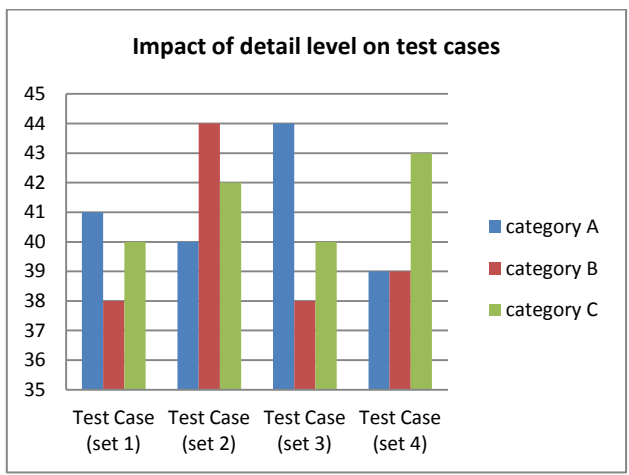

**Fig 4: Impact of detail level**

This study indicates that 29% of the subjects did not read the direction of conveyance of experiment and 15% did not finish the whole layout, so it's impact 44%. This graph illustrates the detail level impact on different test cases. Here are four sets of test cases, in which maximum impact of detail level is 44%. Each set contains three test cases; in the 1st set impact of the detail level on test cases are respectively 41, 38 and 40. In the 2nd set, impact of the detail level is respectively 40, 44 and 42. In the 3$^{rd}$ set, impact is respectively 44, 38 and 40. While in the 4th set impact is respectively 39, 39 and 43.

## D. Not just utilizing the most evident experiment

Amid examination of the few many experiments, it is expected to see the assortment of experiments made. Especially, requested that the analyzers are imaginative in designing legitimate experiments for the framework. Every product framework has an assortment of info conditions, for example, the quantity of complete experiments planned exceptionally awesome, in the testing conditions which consider all mixes of inputs, so outlined test cases by utilizing the strategy which can depict a blend of numerous conditions and create numerous activities correspondingly.

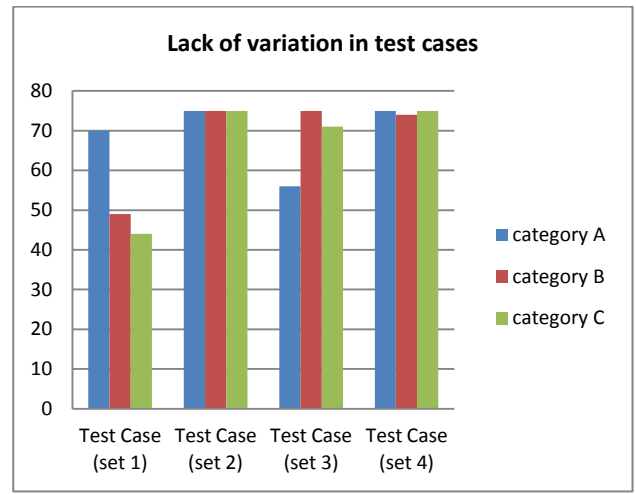

**Fig 5: lack of variation**

This graph demonstrates lack of variation impact on different test cases. Here are four sets of test cases, in which maximum impact of lack of variation is 75%. Each set contains three test cases; in the 1st set impact of the lack of variation on test cases are respectively 70, 49 and 44. In the 2nd set, impact of the lack of variation is respectively 75, 75 and 75. In the 3$^{rd}$ set, impact is respectively 56, 75 and 71. While in the 4th set impact is respectively 75, 74 and 75. Only a few individuals, 25% make any variety inside the framework, or endeavored anything inventive with their experiments. The analyzer often times attempted the limit make a record and the assortment was greatly obliged (generally names and numbers were attempted). The second most fundamental test was looking dated. A couple endeavored to look at a few exchanges which prompted more important tests with marginally higher scope. 75% neglected to endeavor any variety.

## E. Comprehend the motivation behind the framework and current level and connection of testing

Framework effect is vital in numerous viewpoints for comprehension levels and association e.g., detectable quality of the region is possible, what programming thought is used, and how that affect the test approach. Understanding the explanation behind the structure, and current level and association of the system is related to the reflection levels of the system. This is probably the most feathery and hard to handle the thought of a product framework with regards to testing and is by all accounts an understanding that individuals secure after a few years working with the framework.

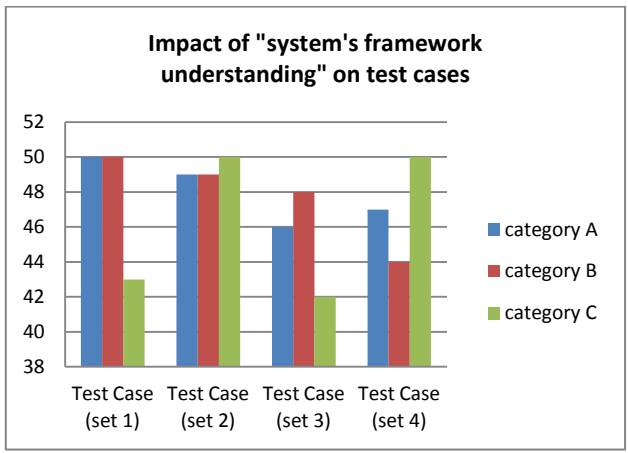

**Fig 6: Impact of system's framework understanding**

Measuring this appreciation is fairly troublesome. It is investigated what number of comprehend that all contributions to the framework were string based, and did not make test cases that is tolerating the test simply handle digits and letters. As much as half of the trials failed on this record (half), which incited a lion's offer of experiments fizzled. This graph shows the system's framework impact on different test cases. Here are four sets of test cases, in which maximum impact of the system's framework is 50%. Each set contains three test cases; in the 1st set impact of the system's framework on test cases are respectively 50, 50 and 43. In the 2nd set, impact of the system's framework is respectively 49, 49 and 50. In the $3^{rd}$ set, impact is respectively 46, 48 and 42. While in the 4th set impact is respectively 47, 44 and 50.

### F. Fix beginning position of the experiment
A common place foul up is to simply depict where the starting position of the test is, i.e., not being specific to the most proficient method to get to the beginning position or which moves must be made before; or simply expecting that a specific area is evident from the experiment connection, or not saying anything by any means.

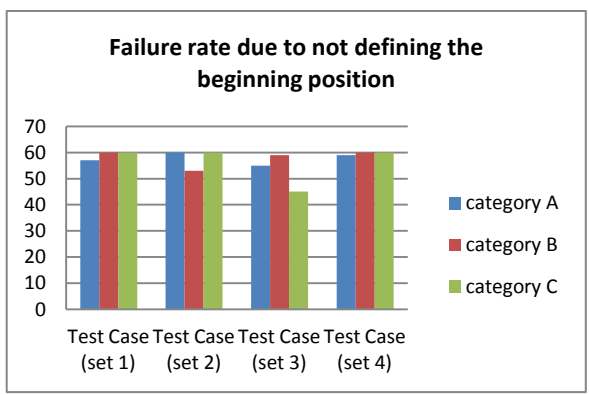

**Fig 7: Impact of defining beginning position**

60% of the test cases fizzled on clarifying an unambiguous beginning position. This graph elaborates failure rate due to not define the beginning position in different test cases. Here are four sets of test cases, in which maximum impact of defining the beginning position is 60%. Each set contains three test cases; in the 1st set impact of the beginning position on test cases are respectively 57, 60 and 60. In the 2nd set, impact of the beginning position is respectively 60, 53 and 60. In the $3^{rd}$ set, impact is respectively 55, 59 and 45. While in the 4th set impact is respectively 59, 60 and 60.

This is obviously effective identified amid mechanization of the experiment that data is lacking to distinguish where the experiment began. The most widely recognized supposition for this action, which targets test of a little program, is tolerating that data, for example, "Begin the system" is sufficient.

### G. Understanding test techniques and how to apply them
This study comprises of the test plan methods and points of interest, cover and variations, inside and out. The first and most clear level is to comprehend what the hypothesis is, and after that the capacity to apply the strategy in the particular case in the framework, which means, finding an area or circumstance where can apply the procedure. Most test outline procedures are identified with information, some are identified by way of execution, and few are identified with request of execution. Likewise blends of methods are conceivable.

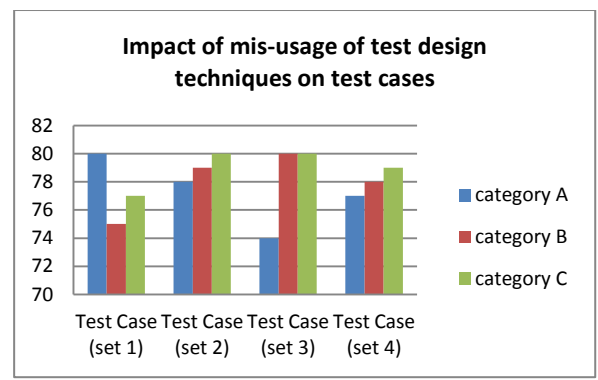

**Fig 8: Impact of miss-usage the techniques**

The positive experiment system (e.g. giving significant information), was the most broadly perceived technique performed, besides the most shocking accomplishment in making an executable trial. Just 20 % honored that for BVA every one of three information is executed, regardless of the fact that was highlighted amid classroom instructing. In this manner, it's impact is 80%. This graph describes miss usage technique's impact on different test cases. Here are four sets of test cases, in which maximum impact of miss usage technique is 80%. Each set contains three test cases; in the 1st set impact of the miss usage technique on test cases are respectively 80, 75 and 77. In the 2nd set, impact of the miss usage technique is respectively 78, 79 and 80. In the $3^{rd}$ set, impact is respectively 74, 80 and 80. While in the 4th set, impact is respectively 77, 78 and 79.

### H. Importance of suppositions
This classification identifies with on which suppositions noticed that an examination has passed or failed. An expert analyzer is inclined to make judgments on exactness and zenith of all parts of the framework. On account of judging a defective necessity, specialists more slanted to expect that the prerequisite is off base and changed. The beginner accepted that what is composed is quite often right, and plan the experiment taking into account this defective supposition.

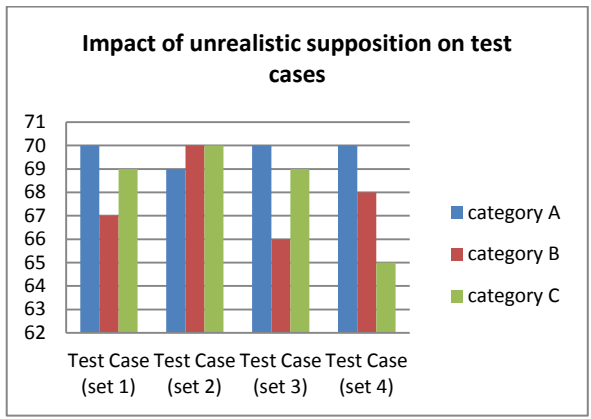

**Fig 9: Impact of unrealistic supposition**

This graph depicts unrealistic supposition impact on different test cases. Here are four sets of test cases, in which maximum impact of unrealistic supposition is 70%. Each set contains three test cases; in the 1st set impact of the unrealistic supposition on test cases are respectively 70, 67 and 69. In the 2nd set, impact of the unrealistic supposition is respectively 69, 70 and 70. In the 3$^{rd}$ set, impact is respectively 70, 66 and 69. While in the 4th set, impact is respectively 70, 68 and 65. More than 70% of analyzers neglected to make a suspicion that coordinated their normal result. Since all analyzers neglected to distinguish what's in store taking into account a comprehension of the framework, they neglected to make reasonable suspicions in connection to that.

## I. Experiment Assessment

The central inspiration driving test execution is to get an estimation of the item quality, by combining a significant course of action of analysis appraisal results. To be a useful trial, it is possible to survey the consequence of the examination to the portrayed criteria. For structures without these criteria the possibility of "Suspicion" is valuable.

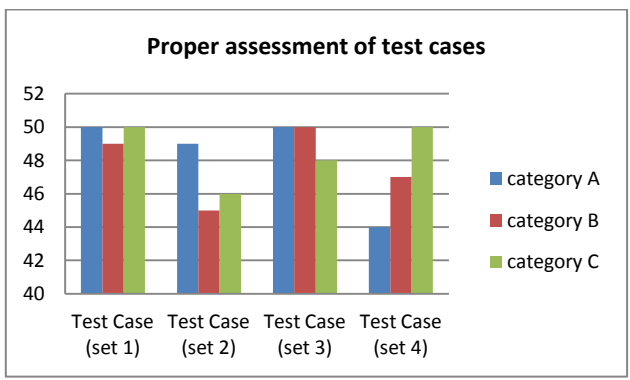

**Fig 10: Proper assessment of test cases**

This graph illustrates the assessment's impact on different test cases. Here are four sets of test cases, in which maximum impact of assessment is 50%. Each set contains three test cases; in the 1st set impact of the assessment of test cases are respectively 50, 49 and 50. In the 2nd set, impact of the assessment is respectively 49, 45 and 46. In the 3$^{rd}$ set, impact is respectively 50, 50 and 48. While in the 4th set impact is respectively 44, 47 and 50. Upwards of 40% missed giving assessment by any means, and around 10% of the experiments not detail an exact assessment that decide the result. In this way, its impact is half, 50%.

## J. Tidy up after an experiment

So also basic to make an accommodating investigation is that it is possible to repeat the examination, again and again. Tidy up after an experiment execution incorporates each one of those activities that are expected to evacuate the impacts of execution to have the capacity to execute once more. In industry, there is ordinarily no extra level of documentation for the test procedure.

This graph demonstrates tidy up's impact on different test cases. Here are four sets of test cases, in which maximum impact of tidy up is 78%. Each set contains three test cases; in the 1st set impact of the tidy up on test cases are respectively 78, 77 and 78. In the 2nd set, impact of the tidy up is respectively 71, 70 and 74. In the 3$^{rd}$ set, impact is respectively 73, 72 and 75. While in the 4th set impact is respectively 78, 72 and 78. 22% of the test cases attempted to clean up. So, its impact is 78%.

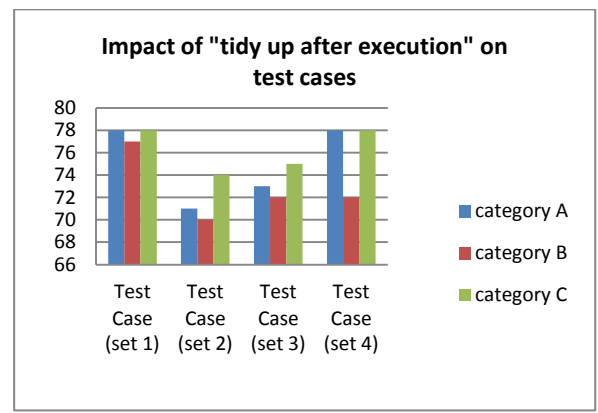

**Fig 11: Impact of tidy up after execution**

This class is effortlessly overlooked, however an undeniable classification while doing mechanization. Tidy up contains numerous activities, and it's especially hard to tidy up in a few frameworks that dependably store information, and don't permit expulsion.

## 4. RESULTS

The Proposed procedure is a great degree gainful that is used to evaluate the way of the item. To test the adequacy of the procedure, diverse exploratory setups are created and the result is dismembered. Inherited Algorithms are definitely not hard to apply to a broad assortment of improvement issues, like the voyaging deals delegate issue, inductive thought learning, booking, and arrangement issues. Programming, testing is like a manner to change issue with the objective that the attempts consumed minimized and the amount of inadequacies perceived extended. Programming, testing is seen as most effort exhausting development in the item change. Regardless of the way that different testing strategies and plenty fullness criteria have been proposed in the written work, in any case it has been watched that no framework/criteria is adequately satisfactory to ensure the movement of weakness free programming huge to the need of change test time to minimize the cost of testing.

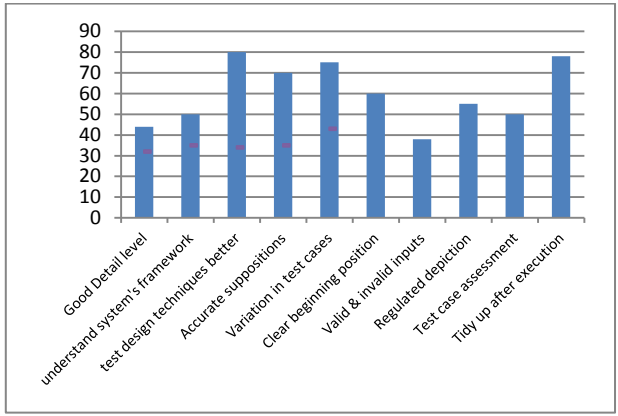

**Fig 12: Factors impact level on test case**

The methodology is additionally time powerful in light of the fact that the estimations are mechanized. The outcome contains least mistakes and exceptionally precise as the information is required. Subsequently, it gives a normal state of testing. The above figure shows the impact level of different factors, having influence on test cases. Impact of good detail level is 44%, understanding system's framework impact is 50%, miss using techniques impact is 80%, suppositions impact is 70%, and variation's impact is 75%, beginning position's impact 60%, input analysis impact is 38%, regulated depiction's impact is 55%, assessment's impact 50% and tidy up impact is 78%. Here are 4 factors that have high impact level that are described in the following table:

**Table 2. Factors having high impact**

| 4 Factors | Impact level |
|---|---|
| Mis-usage of techniques | 80% |
| Tidy up after execution | 78% |
| Lack of Variation | 75% |
| Unrealistic supposition | 70% |

Thus, better enhancement method is presented in the experiment by utilizing Genetic Algorithm. Streamlining strategy obliged distinctive scale undertakings to review. In this study, programming, testing related issues is investigated by utilizing the Genetic Algorithm (GA) approach.

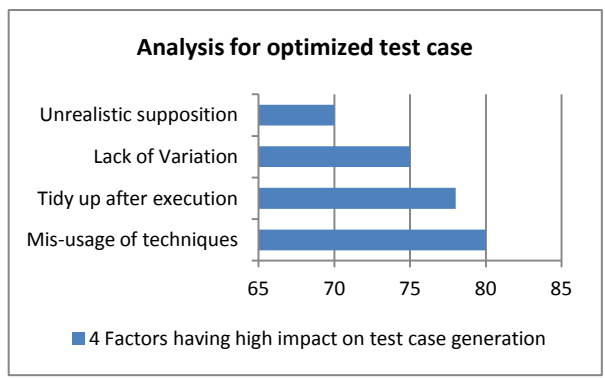

**Fig. 13: Analysis for optimized test case**

This graph is describing the analysis of four factors that are most important for test case optimization. These factors play a significant role and have very high impact on the test case performance. Misusing of test design techniques has higher impact, which is 80%. Other three factors comparatively lower than misusing of test design technique factor.

Optimized test cases can be generated by considering these factors. These aspects are very useful and significant to understand and enhance the performance of test cases.

The composed methodology is satisfactory to the association's configuration measures and gives a smooth stream of data starting with one stage then onto the next. The primary point is to evaluate the achievability of utilizing GA to produce upgraded test information for programming testing. In the wake of applying the some investigation, the better arrangement is created, that is attainable and solid.

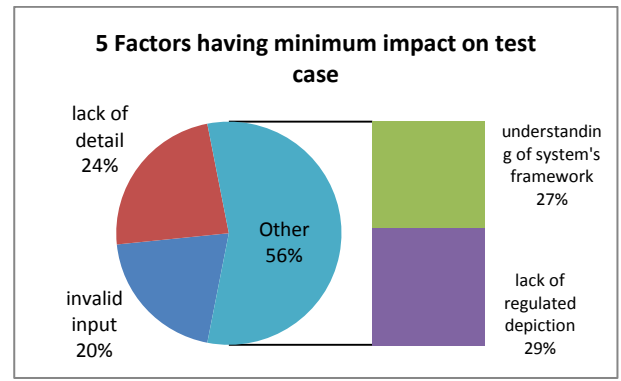

**Fig 14: Minimum impact level on test case**

This graph is an illustration those five factors, that have minimum impact for optimized test case generation. The invalid input factor comparatively lower than other four factors. Lack of regulated depiction factor has 55% impact on test cases, which is higher by comparing other four factors.

So it is hard to test the item altogether, so this procedure is extraordinarily useful in selecting the best game plan of analyses (test suite). The decision relies on upon components that evaluate the analysis, whether it is incredibly or awful.

## 5. CONCLUSION

Testing gives a first means relating to guaranteeing programming execution. The aggregate point of programming industry is really to make certain begin connected with amazing programming for the end client. Be that as it may, one compartment, connected with programming, testing has many fundamental concerns, which are imperative and need to focus on these issues. These issues are compelling era, prioritization of experiments and so forth. These issues are overcome by focusing and core interest. Problems in the product testing zone are typically how to gain an extraordinary appropriate set connected with cases to affirm programming. Some different systems furthermore techniques keep on being proposed relating to transportation consideration of the majority of these issues. The way of the item layout is measured through different procedures and methodologies. If any mix-up happened in any part of the undertaking infers it is essential to change the impacted part of a framework to evacuate the bug. It is conceivable just by recognizing blunders and by measuring the nature of programming. Thusly the way of the item is measured and the result is motorized by the proposed theory of this examination. This rationality gives a capable and proper instrument to evaluate the way of the item. Diverse estimations are proposed closure by the collection strategy for different sorts of vernaculars and the distinctive goof frameworks. Thusly, this investigative work is done successfully with a beneficial methodology proposed in this work.